\documentclass{article}

\usepackage{graphicx,wrapfig}
\usepackage{url}
\graphicspath{ {figures/} }
\usepackage{todonotes}

\usepackage[final,nonatbib]{nips_2017}

\usepackage[utf8]{inputenc} 
\usepackage[T1]{fontenc}    
\usepackage{hyperref}       
\usepackage{url}            
\usepackage{booktabs}       
\usepackage{amsfonts}       
\usepackage{nicefrac}       
\usepackage{microtype}      
\usepackage[para]{footmisc}

\title{Utilizing Domain Knowledge in End-to-End Audio Processing}

\author{
Tycho Max Sylvester Tax\\
Corti, Copenhagen, Denmark\\
\texttt{tt@cortilabs.com}
\And
Jose Luis Diez Antich\\
Audio Analysis Lab,\\
Aalborg University Copenhagen\\
\texttt{jl.diez.antich@gmail.com}
\And
Hendrik Purwins \\
Audio Analysis Lab,\\
Aalborg University Copenhagen\\
\texttt{hpu@create.aau.dk}
\And
Lars Maaløe\\
Corti, Copenhagen, Denmark\\
Technical University of Denmark\\
\texttt{lm@cortilabs.com}
}

\begin{document}

\maketitle

\begin{abstract}

End-to-end neural network based approaches to audio modelling are generally outperformed by models trained on high-level data representations. In this paper we present preliminary work that shows the feasibility of training the first layers of a deep convolutional neural network (CNN) model to learn the commonly-used log-scaled mel-spectrogram transformation. Secondly, we demonstrate that upon initializing the first layers of an end-to-end CNN classifier with the learned transformation, convergence and performance on the ESC-50 environmental sound classification dataset are similar to a CNN-based model trained on the highly pre-processed log-scaled mel-spectrogram features.

\end{abstract}

\section{Introduction}
End-to-end neural network models on image recognition tasks outperform other machine learning approaches by a large margin \cite{NIPS2012_4824} but similarly good results are not seen in the audio domain. Modeling audio is particularly challenging because of long-range temporal dependencies \cite{dieleman2016wavenet} and variations for the same sound due to temporal distortions and phase shifts \cite{ghahremani2016acoustic}. Various papers within automatic speech recognition (ASR), audio classification, and speech synthesis attempt to model audio from raw waveform \cite{dieleman2014end}\cite{dieleman2016wavenet}\cite{hoshen_speech_2015}\cite{collobert2016wav2letter}. Combinations of autoregressive models and dilated convolutions have shown significant improvements over previous results \cite{dieleman2016wavenet}. Still, on tasks such as ASR and environmental sound classification, using traditional transformations, such as (log-scaled mel-)spectrograms or MFCCs generally leads to superior performance.

Modern neural network models are initialized using network-architecture-dependent randomized schemes \cite{DBLP:journals/corr/HeZR015}\cite{pmlr-v9-glorot10a}. In this paper we present preliminary work on initializing a deep neural network for audio classification by explicitly leveraging domain knowledge instead. To do so, we first show that it is possible to train the first layers of a deep neural network model, using unlabelled data, to learn a high-level audio representation. Secondly, we show that upon initializing the first layers of an end-to-end environmental sound classifier with the learned transformation, and keeping the associated parameters fixed during training, convergence and performance are similar to that of a model trained on the high-level representation. This opens up the possibility for training end-to-end neural network models on raw waveform in contexts where there is a limited amount of labeled data available. Finally, we discuss several future directions. It will be particularly interesting to see if fine-tuning of the model after convergence by unfreezing the parameters of the first layers can allow the performance of models trained on raw waveform to surpass those trained on processed features.

\section{Sound classification}
\subsection{Dataset and baseline results}
Several datasets have been generated in recent years to address the previously limited availability of labeled environmental sound data \cite{salamon2014dataset}\cite{45857}\cite{piczak2015esc}. Of particular interest is the Environmental Sound Classification (ESC) dataset, which was released by Piczak in 2015 along with reproducible baseline results for several standard machine learning classifiers (k-NNs, SVMs and RF ensembles) \cite{piczak2015esc} as well as for a convolutional neural network model, which we denote as the PiczakCNN \cite{piczak2015environmental}, and which we use as a baseline for our study. For our analysis, we use the ESC-50 dataset \cite{piczak2015esc}, which contains 2000 5-second long audio clips equally divided among 50 categories. The recordings are pre-arranged in five equally sized folds to facilitate cross-validation.

The PiczakCNN consists of two 2-D CNN layers interleaved with max-pooling followed by two fully connected layers and a softmax output layer. It uses segmented log-scaled mel-spectrograms alongside their deltas as input. The reader is referred to the original paper for more details \cite{piczak2015environmental}. In line with Piczak, we increase the effective number of training samples by making four variations of each ESC-50 clip using time-/pitch-shifting. In addition, each of the mel-spectrograms is cut into segments of 101 frames with 50\% overlap, and silent segments are discarded. At test time, a label is predicted for each of the segments associated with a clip and majority or probability voting is used to determine the final predicted class. Using majority voting the PiczakCNN model achieves a $\sim$62\% accuracy on the test set averaged over the five folds \cite{piczak2015environmental}.

Since we are interested in an end-to-end approach we add several 1-D CNN layers to the PiczakCNN architecture to allow raw waveform as input\footnote{Source code for the project will be made available at: \url{https://github.com/corticph/MSTmodel}.}. These layers by themselves form the model that is trained separately to learn the mel-spectrogram transformation after which the learned weights are used to initialize the end-to-end classifier and kept frozen for the classification task.

For this study, we first replicate Piczak's results using mel-spectrograms and their deltas to ensure that our set-up performs adequately. As we attempt to learn the mel-spectrogram transformation only, and not also the associated deltas, we benchmark our results on raw speech against PiczakCNN performance without deltas, which is $\sim$53\% on the test set using majority voting, and averaged over the 5 folds. 

\subsection{Learning the mel-spectrogram transformation}
\subsubsection{Experimental set-up}
After exploring several architectures for the mel-spectrogram transformation model (MSTmodel) we picked the architecture summarized in Table \ref{table:network}. The reason is twofold: firstly, it shows excellent performance on the modelling task, and secondly, its parameters are matched in terms of kernel size and stride to the window-and hop-size of the short-time Fourier transform (STFT) step in the mel-spectrum calculation. The second and third layer are qualitatively similar to those used for temporal modelling in \cite{collobert2016wav2letter}. 'SAME' padding is used in the MSTmodel so that the output feature maps are of the same spatial dimension as the input feature maps in order not to lose information at the edges of the signal during the forward pass.

\begin{table}[h!]
\centering
\caption{Details of the MSTmodel.}
\label{table:network}
\begin{tabular}{@{}llllll@{}}
\toprule
 Layer type  & Number of filters & Filter size & Stride & Activation function & Padding \\ \midrule
1D convolution & 512 & 1024  & 512   & ReLU & SAME   \\
1D convolution & 256 & 3  & 1   & ReLU & SAME   \\
1D convolution & 60 & 3  & 1   & Tanh & SAME  \\ \bottomrule
\end{tabular}
\end{table}

The MSTmodel is trained stand-alone, in supervised fashion, by taking raw audio clips as input and their corresponding mel-spectrograms as labels. The raw waveform recordings are re-sampled at 22050Hz from their original 44.1kHz, and their amplitudes are normalized per segment (corresponding to a single mel-spectrogram of framelength 101) between -1 and 1 through division with the maximum absolute value of the segment. A target mel-spectrogram is generated using the Librosa package \cite{mcfee_brian_2017_293021} by applying a mel-filterbank to the magnitude spectrum of each segment (window size = 1024, hop size = 512, and mel bands = 60), and then taking its logarithm. The mel-spectrogram values are normalized and re-scaled between -1 and 1 using trainset statistics.

To train the MSTmodel we use a simple mean squared error (MSE) loss between the predicted representations and the target mel-spectrograms, where the last frame of each label is sliced off to match dimensions\footnote{Alternatively, one could pad the input data but the current approach led to sufficiently accurate mel-spectrograms for our purposes.}. We use Adam \cite{kingma2014adam} with a constant learning rate ($\mathrm{3e^{-4}}$) and a batch size of 100 for optimization, and prevent overfitting through early stopping based on the performance on the pre-allocated validation set. Separate models are trained for each pre-assigned fold to ensure that upon initializing the end-to-end classifier, it has not previously 'seen' test or validation data. 

\subsubsection{Results and discussion}
\begin{figure}[h]
  \centering
  \includegraphics[width=0.9\textwidth]{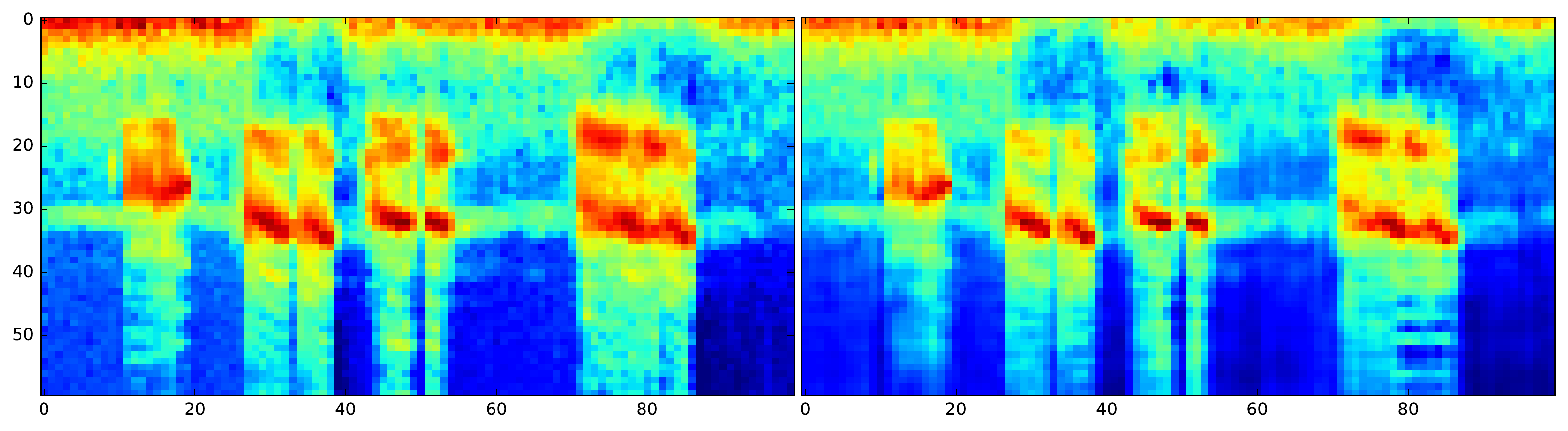}
\caption{Comparison of a mel-spectrogram (left) and its prediction (right). The signal is randomly chosen from the test set.}
  \label{fig:learned_spectrogram}
\end{figure}

Figure \ref{fig:learned_spectrogram} shows the learned mel-spectrogram (right) and its target (left) for a randomly selected example from the test set. The prediction is visually similar to the target although the spectrum seems slightly smoothed. We have verified the learned transformation is not domain dependent by comparing the model predictions with their targets on pure tones, speech, and music.

\begin{figure}[h]
\centering
\begin{minipage}{.48\textwidth}
\vspace*{-2cm}
  \centering
  \includegraphics[width=\linewidth]{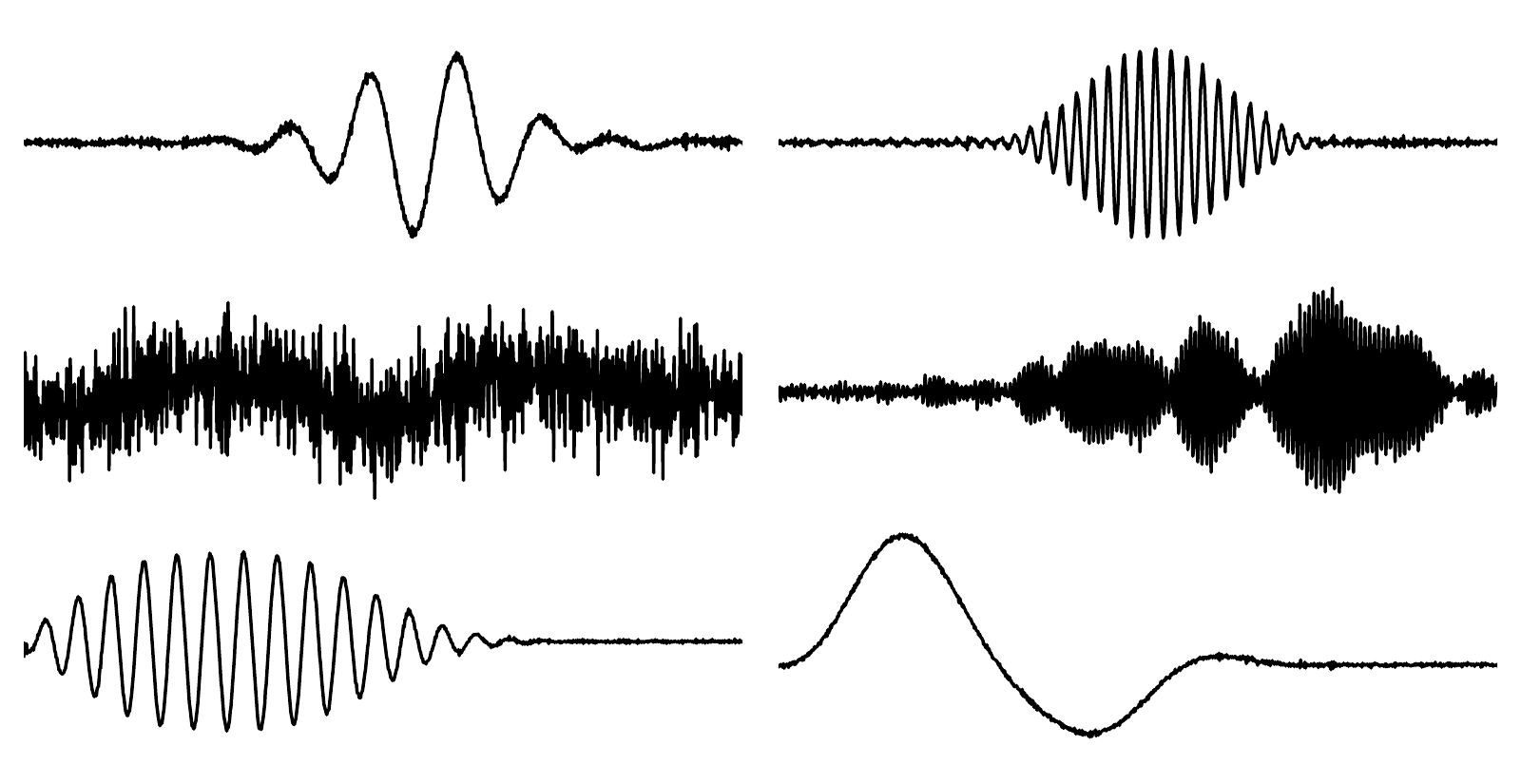}
  \caption{Subset of the filters learned by the first MSTmodel layer.}
  \label{fig:filters}
\end{minipage}%
\hfill
\begin{minipage}{.45\textwidth}
  \centering
  \includegraphics[width=\linewidth]{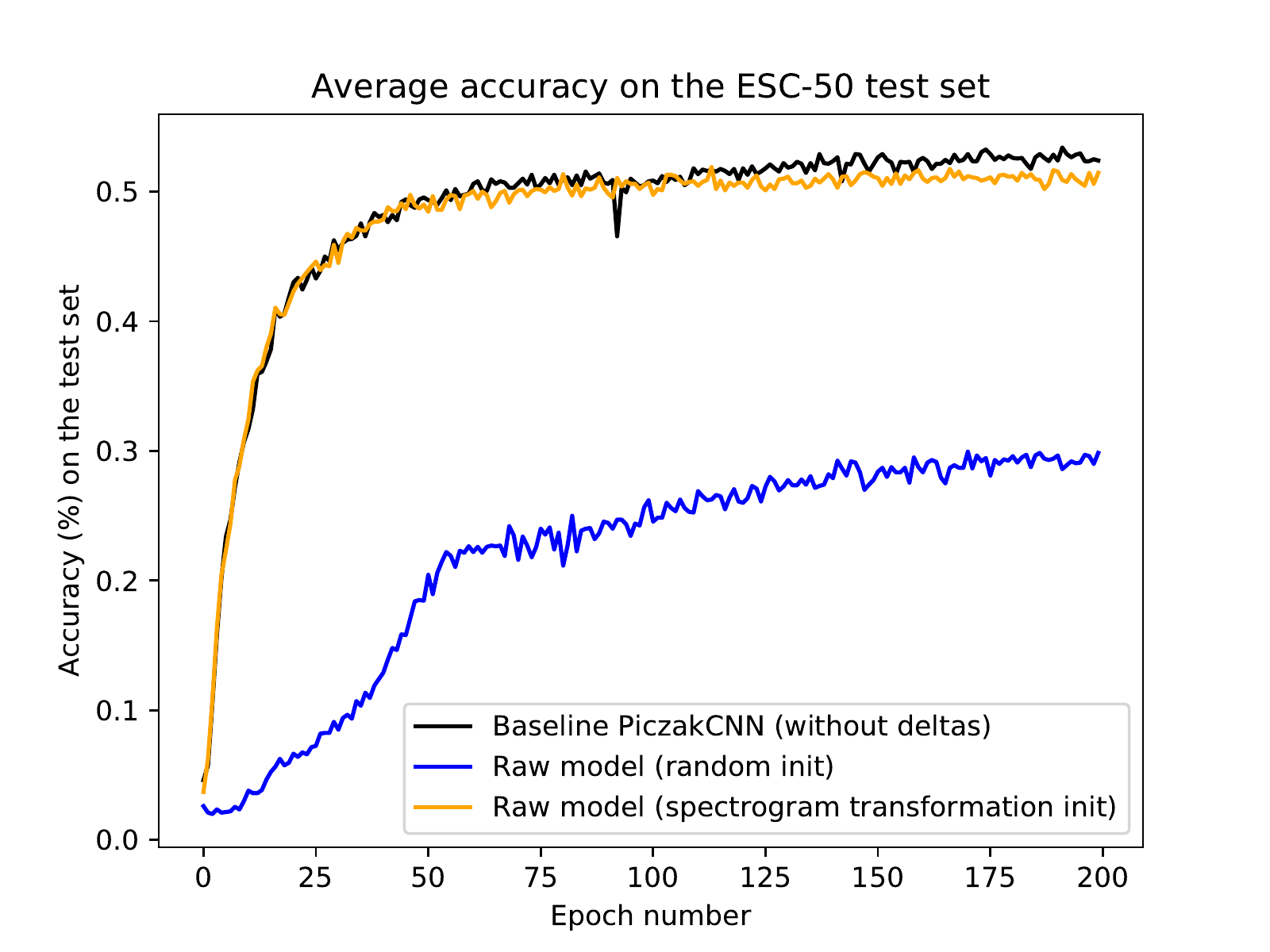}
  \caption{Test accuracy averaged over the folds for: (a) baseline PiczakCNN, (b) network with random initialization (Xavier), and (c) network initialized with the spectrogram transformation.}
  \label{fig:accuracy}
\end{minipage}
\end{figure}

In Figure \ref{fig:filters}, we plot some of the learned filters of the first layer of the MSTmodel. It can be seen that our network learns frequency decompositions such as wavelets and band-pass filters that are qualitatively similar to those reported in previous studies \cite{tuske2014acoustic}\cite{aytar2016soundnet}\cite{hoshen_speech_2015}. It is interesting that our network discovers the same representations despite being trained on a different task.

\subsection{End-to-end environmental sound classification}
\subsubsection{Experimental set-up}
We assess the success of learning the mel-spectrogram transformation by performance on the ESC-50 dataset. To do so, we train three models, where in each case we adhere to the pre-defined cross-validation structure for the ESC-50 dataset \cite{piczak2015environmental}: 1. the baseline PiczakCNN model on mel-spectrograms without deltas, 2. the PiczakCNN model with the three-layer MSTmodel architecture added, and with random initialization using the Xavier scheme \cite{pmlr-v9-glorot10a}, and 3. the same model as in 2 but with the pre-trained MSTmodel layers. In the second classifier, dropout layers (keep probability = 0.5) are added to the MSTmodel-layers after each of the non-linearities to prevent overfitting. For the third classifier we keep the MSTmodel parameters frozen to the learned mel-spectrogram transformation during training of the deeper layers. 

The training scheme used is largely the same as the one proposed by Piczak \cite{piczak2015environmental} with minor adaptions to the hyperparameter choices and normalization---after normalization the mel-spectrogram values are re-scaled between -1 and 1 based on the minimum and maximum of the trainset. We use a cross-entropy loss function and stochastic gradient descent with Nesterov momentum (0.9), a batch size of 500, and a learning rate of $\mathrm{5e^{-3}}$, and we train the models for 200 epochs. The originally proposed L2-weight regularization is not used in the final experiments since it did not improve performance. At test-time, majority voting is used to determine the class of the test sample based on all its associated overlapping segments.

\subsubsection{Results and discussion}
The performance of the three models outlined above is presented in Figure \ref{fig:accuracy} on the test set averaged over the five folds. We see that initializing the weights of the first layers with the learned mel-spectrogram transformation, and keeping them fixed throughout training, results in better convergence and increases performance on raw waveform approximately to mel-spectrogram-levels.

When using neural networks for audio modelling, architectural choices are sometimes made that appear inspired by deterministic feature extracting methods. For instance, max-pooling is often used to perform a summarizing process comparable to the mel-filter bank operating on a mel-spectrogram. In addition, log-layers are sometimes used to compress the learned internal representation \cite{sainath2015learning}\cite{hoshen_speech_2015}. CNNs are highly flexible models that can approximate complex mappings. By leveraging this capability in our approach, and forcing the network to learn such transformations implicitly, we limit the need for ad-hoc architectural choices.

\section{Conclusion and future directions}
This proof-of-concept study shows that 1) through supervised training, a simple CNN architecture can learn the log-scaled mel-spectrogram transformation from raw waveform and 2) that initializing an end-to-end neural network classifier with the learned transformation yields a performance comparable to a model trained on the highly processed mel-spectrograms. These findings show that incorporating knowledge from established audio signal processing methods can improve performance of neural network based approaches on audio modeling tasks. This preliminary work opens up the possibility for a myriad of follow-up experiments. Most notably, it will be interesting to fine-tune the previously fixed parameters of the first layers of the classifier to determine if different representations are learned and whether they are more informative than mel-spectrograms. If so, the robustness of these representations can be further increased through an abundance of available unlabelled audio data. This parallels work by Jaitly and Hinton who use generative models to leverage unlabeled data to learn robust features \cite{jaitly2011learning}.

\paragraph{Acknowledgments}
The authors would like to thank Karol Piczak for making the code to reproduce his baseline results publicly available, and for answering several questions relating the ESC dataset. In addition, we would like to thank the Corti team, Lasse Borgholt and Alexander Wahl-Rasmussen in particular, for insightful feedback, and proofreading the manuscript.
\newpage

\bibliographystyle{abbrv}
\bibliography{bibliography}

\end{document}